\documentclass[a4paper]{article}

\usepackage{INTERSPEECH2022}
\usepackage{hyperref}
\usepackage{makecell}
\usepackage{multirow}
\usepackage{enumitem}

\makeatletter
\def\blfootnote{\xdef\@thefnmark{}\@footnotetext}
\makeatother

\title{An Empirical Study of Language Model Integration for Transducer based Speech Recognition}
\name{Huahuan Zheng$^1$, Keyu An$^1$, Zhijian Ou$^1$$^{\dagger}$\thanks{$\dagger$ Corresponding author, also affiliated with Beijing National Research Center for Information Science and Technology, China. This work is supported by NSFC 61976122 and Tsinghua University - Meituan Joint Institute for Digital Life.}, Chen Huang$^2$, Ke Ding$^2$, Guanglu Wan$^2$}
\address{
  $^1$Speech Processing and Machine Intelligence (SPMI) Lab, Tsinghua University, China\\$^2$Meituan, China}
\email{\{zhh20,aky19\}@mails.tsinghua.edu.cn, ozj@tsinghua.edu.cn, \{huangchen09,dingke02,wanguanglu\}@meituan.com}

\begin{document}

\maketitle
\begin{abstract}
Utilizing text-only data with an external language model (ELM) in end-to-end RNN-Transducer (RNN-T) for speech recognition is challenging.
Recently, a class of methods such as density ratio (DR) and internal language model estimation (ILME) have been developed, outperforming the classic shallow fusion (SF) method. 
The basic idea behind these methods is that RNN-T posterior should first subtract the implicitly learned internal language model (ILM) prior, in order to integrate the ELM.
While recent studies suggest that RNN-T only learns some low-order language model information, the DR method uses a well-trained neural language model with full context, which may be inappropriate for the estimation of ILM and deteriorate the integration performance. Based on the DR method, we propose a low-order density ratio method (LODR) by replacing the estimation with a low-order weak language model.
Extensive empirical experiments are conducted on both in-domain and cross-domain scenarios on English LibriSpeech \& Tedlium-2 and Chinese WenetSpeech \& AISHELL-1 datasets. 
It is shown that LODR consistently outperforms SF in all tasks, while performing generally close to ILME and better than DR in most tests.
\end{abstract}

\section{Introduction}
\label{sec:intro}
A trend of speech recognition research is moving from hybrid models to end-to-end (E2E) models such as connectionist temporal classification (CTC) \cite{graves2006connectionist}, recurrent neural network transducer (RNN-T) \cite{graves2012sequence} and attention-based encoder-decoder (AED) \cite{chorowski2015attention}. E2E models have been shown to outperform hybrid ones in in-domain tasks \cite{he2019streaming, gulati2020conformer, li2020developing, li2020comparison}, when trained on large amounts of paired speech-text data.

However, in practice, text-only data is easily available, allowing improving the performance of ASR models in lower cost, compared to increasing the paired labeled speech-text data. Language model (LM) integration is a widely-used approach to boost ASR performance with text-only data for both in-domain test and cross-domain adaptation.
ASR decoding amounts to maximizing the posterior probability $P(Y|X)$, where $Y$ is token sequence represented by words, sub-words or characters, and $X$ is the observed speech data.
Since a hybrid system explicitly learns the acoustic model (AM), integrating the external language model (ELM) is straightforward, by following the Bayes rule:
\vspace{-1mm}
\begin{equation}
  \label{eq:hybrid lm integration}
  \hat{Y} = \mathop{\arg\max}_{Y}\left[ P_{\text{AM}}(X|Y)P_{\text{ELM}}(Y) \right]
  \vspace{-1mm}
\end{equation}
However, E2E systems like RNN-T jointly learn the posterior probability $P(Y|X)$ in a unified model, with no separation of AM and LM, making the LM integration for E2E models not as easy as that of hybrid models.

A number of ``fusion'' approaches have been proposed to address the problem of LM integration for E2E models, such as shallow fusion \cite{mikolov2010recurrent, hannun2014deep, kannan2018analysis}, deep fusion \cite{gulcehre2015using} and cold fusion \cite{sriram2017cold}, among which the shallow fusion is the most popular one.
SF just conducts log-linear interpolation between the scores of the E2E model and the ELM. Recently, a class of internal language model (ILM) estimation methods have been developed, outperforming SF, which were initially studied for RNN-T \cite{variani2020hybrid} and later extended to AED \cite{meng2021internal,zeineldeen2021investigating}. In this paper, we constrain our study of ILM to RNN-T.

The basic idea of ILM is that an RNN-T model $P_{\text{RNN-T}}(Y|X)$, through training over labeled data, has captured LM information from the transcripts, and such information is thought to be encapsulated in the so-called ILM.
Thus, if the ILM from the RNN-T can be estimated in some way, the LM integration for the RNN-T model can be achieved by subtracting the ILM and integrating the ELM in the manner of Eq. \eqref{eq:hybrid lm integration} for hybrid models:
\begin{equation}
  \label{eq:e2e lm integration via subs ilm}
  \hat{Y} = \mathop{\arg\max}_{Y}\left[ \frac{P_{\text{RNN-T}}(Y|X)}{P_{\text{ILM}}(Y)}P_{\text{ELM}}(Y) \right]
\end{equation}

Based on the above idea, many ILM estimation methods are proposed, differing mainly in the manner how the ILM is estimated.
In hybrid autoregressive transducer (HAT) \cite{variani2020hybrid}, the distributions for the blank label \verb|<blk>| and normal labels are re-defined via a Sigmoid function and a Softmax function separately.
\cite{meng2021internal} proposes the ``internal lanugage model estimation'' (ILME) method, which zeros out the acoustic hidden state and normalizes the logits from the joint network just over non-blank labels.
In a more complex method, called mini-LSTM \cite{zhou2021language}, the acoustic hidden state is calculated via an additional LSTM.
In these methods, the ILM probability is the product of the resulting label probabilities, which basically still come from the RNN-T model.
Instead, the density ratio (DR) method \cite{mcdermott2019density} uses a neural LM separately trained over the E2E training transcripts as an estimate of ILM, whose structure and size are usually set to match the prediction network in RNN-T.

As reported in \cite{variani2020hybrid, meng2021internal}, the perplexities (PPLs) of the estimated ILMs during RNN-T training first decrease and then increasingly converge to 
rather larger PPLs than those of the well-learned neural LMs used in the DR method (e.g., 99.4 and 30.1 in \cite{meng2021internal}).
Moreover, it is found in \cite{variani2020hybrid, ghodsi2020rnn} that RNN-T using only one history label for prediction network can achieve competitive performance to the one with full context, which also suggests RNN-T only learns some low-order language information, but not a strong LM.
Inspired by these previous findings, we propose to train a low-order weak LM (such as a bi-gram LM) over the transcripts\footnote{When the transcripts are not available, training a low-order LM over some other text corpus also produces competitive performance, as shown later in our experiments.}, use it as the estimate of ILM, and substitute to Eq. (\ref{eq:e2e lm integration via subs ilm}) to perform LM integration.
The proposed method, which we refer to as low-order density ratio (LODR), can be viewed as an extension of the DR method, estimating the ILM with low-order information, instead of training well-learned neural LMs as in standard DR method.

Extensive experiments are conducted in this paper to compare the performances of SF, DR, ILME, and LODR on RNN-T, evaluated on both in-domain and cross-domain tasks with text-only data.
It is shown in this study that ILME and LODR performs close to each other, both better than SF (in all tests) and DR (in most tests).
This further consolidates and deepens our understanding of the ILM in RNN-T, and would be helpful for future work to further advance LM integration for RNN-T.
The code will be released upon the acceptance of the paper.

All comparisons are made in decode-time manner, thus the works like \cite{variani2020hybrid, zhou2021language, meng2021ilmt, meng2021minimum} that modify the architecture or training objective of RNN-T are not included in this work.
It should also be clarified that, when discussing ``cross-domain", we assume the source domain and target domain are matched in acoustics, otherwise Eq. \eqref{eq:e2e lm integration via subs ilm} cannot be adopted.
\begin{table*}[th]
  \caption{Performance of LM integration methods, represented as word error rate (WER \%) on LibriSpeech and character error rate (CER \%) on WenetSpeech. The perplexity (PPL) of the ILM is computed on the transcript of each dataset, which could roughly measure the similarity of the ILM and the transcript corpus distributions. $\lambda_0$ is the weight of the ILM in each method. For all methods, $\lambda_1$ denotes the weight of the shared ELM and $\beta$ denotes the length reward. ``avg.'' is the average WER/CER on all dev and test datasets. ``Rel \%'' measures the relative reduction of WER (CER) compared to ``No LM'' setup.}
  \label{tab:in-domain}
  \vspace{-2mm}
  \centering
  \begin{tabular}{|c|c|c|c|c||c|c|c|c|c||c|}
    \hline
    \multirow{3}{*}{Method} & \multirow{3}{*}{ILM PPL} & \multirow{3}{*}{$\lambda_0$} & \multirow{3}{*}{$\lambda_1$} & \multirow{3}{*}{$\beta$} & \multicolumn{6}{c|}{LibriSpeech}                                                                                                                        \\
    \cline{6-11}
                            &                          &                              &                              &                          & \multicolumn{2}{c|}{dev}                  & \multicolumn{2}{c|}{test} & \multirow{2}{*}{avg.} & \multirow{2}{*}{Rel \%}                                 \\
    \cline{6-7} \cline{8-9}
                            &                          &                              &                              &                          & clean                                     & other                     & clean                 & other                   &               &               \\
    \hline
    No LM                   & -                        & -                            & -                            & -                        & 2.18                                      & 5.33                      & 2.40                  & 5.42                    & 3.81          & -             \\
    \hline
    SF                      & -                        & -                            & 0.625                        & 1.0                      & 1.82                                      & 4.06                      & 1.96                  & 4.42                    & 3.04          & 20.2          \\
    \hline
    DR                      & 24.72                    & -0.125                       & 0.75                         & 0.5                      & 1.79                                      & 4.00                      & 1.97                  & 4.31                    & 3.00          & 21.3          \\
    \hline
    ILME                    & 50.21                    & -0.125                       & 0.75                         & 1.0                      & 1.78                                      & 3.99                      & 1.92                  & 4.35                    & \textbf{2.99} & \textbf{21.5} \\
    \hline
    LODR                    & 100.94                   & -0.125                       & 0.75                         & 0.75                     & 1.83                                      & 4.00                      & 1.94                  & 4.34                    & 3.01          & 21.0          \\

    \hline
    \hline
    \multirow{3}{*}{Method} & \multirow{3}{*}{ILM PPL} & \multirow{3}{*}{$\lambda_0$} & \multirow{3}{*}{$\lambda_1$} & \multirow{3}{*}{$\beta$} & \multicolumn{6}{c|}{WenetSpeech}                                                                                                                        \\
    \cline{6-11}
                            &                          &                              &                              &                          & \multicolumn{2}{c|}{\multirow{2}{*}{dev}} & \multicolumn{2}{c|}{test} & \multirow{2}{*}{avg.} & \multirow{2}{*}{Rel \%}                                 \\
    \cline{8-9}
                            &                          &                              &                              &                          & \multicolumn{2}{c|}{}                     & net                       & meeting               &                         &                               \\
    \hline
    No LM                   & -                        & -                            & -                            & -                        & \multicolumn{2}{c|}{11.14}                & 12.75                     & 20.88                 & 14.05                   & -                             \\
    \hline
    SF                      & -                        & -                            & 0.25                         & 3.125                    & \multicolumn{2}{c|}{9.19}                 & 11.73                     & 18.36                 & 12.37                   & 12.0                          \\
    \hline
    DR                      & 37.89                    & 0.0                          & 0.25                         & 3.125                    & \multicolumn{2}{c|}{9.19}                 & 11.73                     & 18.36                 & 12.37                   & 12.0                          \\
    \hline
    ILME                    & 94.32                    & -0.125                       & 0.375                        & 3.0                      & \multicolumn{2}{c|}{9.10}                 & 11.56                     & 18.26                 & 12.25                   & 12.8                          \\
    \hline
    LODR                    & 79.33                    & -0.125                       & 0.375                        & 3.125                    & \multicolumn{2}{c|}{9.07}                 & 11.54                     & 18.23                 & \textbf{12.22 }         & \textbf{13.0}                 \\
    \hline
  \end{tabular}

  \vspace{-3mm}
\end{table*}
\section{LM Integration Methods for RNN-T}

\subsection{RNN Transducer}
\label{sec:RNN-T}
The RNN-T model \cite{graves2012sequence} consists of an acoustic encoder (a.k.a. the transcription network), a prediction network (PN) and a joint network. With given acoustic features $X=(\mathbf{x}_1, ..., \mathbf{x}_T)$ and the token sequence $Y=(y_0, ..., y_U)$, the posterior probability is defined as:
\vspace{-1.5mm}
\begin{equation}
  \label{eq:RNN-T p_y_x}
  \begin{aligned}
    P_{\text{RNN-T}}(Y|X) & = \sum_{\tilde{Y}\in \mathcal{B}^{-1}(Y)} P(\tilde{Y}|X)                             \\
           & =\sum_{\tilde{Y}\in \mathcal{B}^{-1}(Y)} \prod_{i=1}^{T+U} P(\tilde{y}_i|X, y_{0:u})
  \end{aligned}
  \vspace{-1mm}
\end{equation}
\begin{equation}
  \label{eq:rnn-t joint}
  P(\tilde{y}_i|X, y_{0:u}) = \text{Softmax}
  \left[J\left(\mathbf{g}_u, \mathbf{f}_t\right)\right]
\end{equation}
Here $\tilde{Y}=(\tilde{y}_1,...,\tilde{y}_{T+U})$ represent the alignment sequence including the blank label \verb|<blk>|.
$t\in [1, T]$ and $u\in [0, U]$ denote the number of speech frames and that of non-blank labels up to emitting $\tilde{y}_i$. $\mathcal{B}(\cdot)$ represents the alignment mapping of $\tilde{Y}$ to $Y$ by removing \verb|<blk>|. The token-level probabilities are computed by the joint network as in Eq.~\eqref{eq:rnn-t joint}, where $\mathbf{g}_u$ and $\mathbf{f}_t$ denotes the output hidden features of the PN and encoder.  $J(\cdot)$ denotes the joint network, which consists of linear layers and non-linear activations in the common RNN-T architecture.

\vspace{-1mm}
\subsection{Shallow Fusion}

The shallow fusion (SF) method takes the linear combination of log scores of E2E models and the ELMs, with one extra parameter $\beta$ as reward for the sequence length $|Y|$. The hypothesis $\hat{Y}$ is selected as follows:
\vspace{-2mm}
\begin{equation}
  \label{eq:shallow fusion}
  \hat{Y} = \mathop{\arg\max}_{Y} \left[\log P_{\text{E2E}}(Y|X) + \lambda \log P_{\text{ELM}} (Y) + \beta |Y|\right]
\end{equation}
%

\vspace{-2mm}
\subsection{Density Ratio}

The density ratio (DR) method \cite{mcdermott2019density} is proposed as an extension of SF, and makes the assumption that the model distribution $P(Y|X)$ captured by RNN-T can be factorized into acoustic and linguistic parts as in the hybrid system:
\vspace{-1mm}
\begin{equation}
  P_{\text{RNN-T}}(X|Y) = P_{\text{RNN-T}}(X)\frac{P_{\text{RNN-T}}(Y|X)}{P_{\text{RNN-T}}(Y)}
  \vspace{-1mm}
\end{equation}
where $P_{\text{RNN-T}}(X)$ models the marginal distribution of data. The DR method further assumes: (1) the acoustic models of the source domain (where the RNN-T is trained) and the target domain are consistent, both modeled by $P_{\text{RNN-T}}(X|Y)$; (2) the linguistic distribution for the two domains can be separately estimated via LMs $P_{ILM}(Y) \approx P_{\text{RNN-T}}(Y)$ and $P_{ELM}(Y)$, respectively.
According to \cite{mcdermott2019density}, $P_{ILM}(Y)$ is estimated by a separately trained neural LM over source domain transcripts.
$P_{ELM}(Y)$ can be trained over the target domain corpus.
With these assumptions, decoding for recognizing utterances from the target domain can be derived the same as shown in Eq. (\ref{eq:e2e lm integration via subs ilm}).
Introducing LM weights $(\lambda_0, \lambda_1)$ and length reward $\beta$, we have
%
\begin{equation}
  \label{eq:density ratio hypo select}
  \begin{aligned}
    \hat{Y} = \mathop{\arg\max}_{Y} & \left[\log P_{\text{RNN-T}}(Y|X)+\lambda_0 \log P_{\text{ILM}}(Y)\right. \\
                                    & \left. + \lambda_1 \log P_{\text{ELM}}(Y) + \beta|Y|\right]
  \end{aligned}
\end{equation}
When $\lambda_0=-1, \lambda_1=1$ and $\beta=0$, Eq.~\eqref{eq:density ratio hypo select} is equivalent to Eq.~\eqref{eq:e2e lm integration via subs ilm}, in a strict manner of Bayes rule. 

\vspace{-2mm}
\subsection{ILME}

Inspired by \cite{variani2020hybrid}, ILME \cite{meng2021internal} applies Proposition 1 in \cite[Appendix.~A]{variani2020hybrid} to estimate the ILM via zeroing out the acoustic part in Eq.~\eqref{eq:rnn-t joint}. The proposition claims that, if $J\left(\mathbf{g}_u, \mathbf{f}_t\right)\approx J\left(\mathbf{g}_u, \mathbf{0}\right) + J\left(\mathbf{0}, \mathbf{f}_t\right)$ is satisfied, we have
\vspace{-1.5mm}
\begin{equation}
  P_{\text{RNN-T}}(y_{u+1}|y_{0:u}) \propto \exp \left( J(\mathbf{g}_u)\right)
  \vspace{-1.5mm}
\end{equation}
where we omit the $\mathbf{0}$ to simplify the notations.
So the ILM of RNN-T can be computed by applying Softmax to the token logits excluding \verb|<blk>|, i.e.
\vspace{-2mm}
\begin{equation}
  P_{\text{ILM}}(Y) = \prod_{u=0}^U \text{Softmax}
  \left[ J_{\setminus\verb|<blk>|}(\mathbf{g}_u) \right]
  \vspace{-2mm}
\end{equation}
The ILME decoding basically follows Eq.~\eqref{eq:density ratio hypo select}, except that the ILM is estimated from RNN-T itself.
It is reported that the simple joint network described in Sec.~\ref{sec:RNN-T} makes RNN-T possible to roughly satisfy the condition of the proposition \cite{variani2020hybrid}. 

\section{Low-order Density Ratio Estimate}
\subsection{Design of LODR}
\label{sec:design of low-order ilm}
As discussed in Sec.~\ref{sec:intro}, previous works have shown that the ILM of RNN-T actually captures small amount of LM information from the transcripts \cite{variani2020hybrid, meng2021internal}, and RNN-T only makes use of limited context and low-order information in the prediction network \cite{variani2020hybrid, ghodsi2020rnn}.
In contrast, the DR method uses a well-learned LM with full context as the estimation of ILM.
Inspired by the findings, we hypothesize that the original DR setting is inappropriate and may deteriorate the performance of integration.

To investigate the performance of DR using a low-order weak ILM, we separately train a low-order LM on the transcripts and follow the Eq.~\eqref{eq:density ratio hypo select} to do the integration. In implementation, we train a bi-gram LM on the training transcripts, and prune the bi-grams except for the most frequent 20k ones. This would produce a very small model, typically around 250 kB on disk. Note that if the modeling units are in small granularity like alphabets, the number of bi-grams is probably not enough up to 20k; in that case, one may need to use relatively higher order of n-gram. In our experiments, we use 1024 word-pieces for English and around 5k characters for Chinese as the modeling units, where the numbers of bi-grams are both more than 100k. We leave further discussion to the Sec.~\ref{sec:discussion of LODR}.

\vspace{-2mm}
\subsection{Implementation of Score Interpolation}
\label{sec:impl of lm interpolation}
\vspace{-0.5mm}
As is shown in Eq.~\eqref{eq:shallow fusion} and Eq.~\eqref{eq:density ratio hypo select}, SF, DR, ILME and LODR are all essentially conducting linear interpolation between scores of the RNN-T and the LMs, where the hyperparameters, i.e., the $\lambda_0, \lambda_1$ and $\beta$, are shown to be important to the integration performance \cite{meng2021internal, zhou2021language, mcdermott2019density}.
Since lacking of common hyperparameter tuning setup in the literature and grid-search becomes too expensive as the number of hyperparameters exceeds two, in this work, we detail our hyperparameter searching method, which is coordinate descent \cite{wright2015coordinate} combined with binary search:
\vspace{-1mm}
\setlist{leftmargin=5mm}
\begin{enumerate}
  \setlength{\itemsep}{0pt}
  \setlength{\parsep}{0pt}
  \setlength{\parskip}{0pt}
  \item Initialize the searching ranges and minimum interval sizes for each hyperparameter.
  \item Tune one hyperparameter per iteration. In an iteration, fix other hyperparameters and only search for one. Binary search is used for tuning one hyperparameter, and is done when the searching range is smaller than the minimum interval size.
  \item Continue the loop until there is not any better combination. Once the tuned hyperparameters are at the boundary of the initial range, we extend the range and continue the loop.
\end{enumerate}
\vspace{-1mm}
The hyperparameters are first tuned on a held-out validation set, then evaluated on the test sets.
In the experiments, all LM integration methods follow the same steps to tune their hyperparameters, except that there are two hyperparameters for SF as Eq. \eqref{eq:shallow fusion}, while three for DR, ILME and LODR as Eq.~\eqref{eq:density ratio hypo select}.

\vspace{-2mm}
\section{Experiment}

Experiments are taken on 1000-hour WenetSpeech \cite{zhang2021wenetspeech} Chinese dataset\footnote{The full dataset includes 10k hours of labeled data. We just take the 1000-hour train-M subset.},
960-hour LibriSpeech \cite{panayotov2015librispeech} English dataset and an in-house Chinese dataset \textit{Tasi} of around 4.5k hours. The 200M-char WenetSpeech corpus and 800M-word LibriSpeech corpus  are taken for in-domain test. For cross-domain scenarios, we use Tedlium-2 \cite{rousseau2014enhancing}, AISHELL-1 \cite{bu2017aishell} and an in-house TV-news dataset to evaluate on the three training sets respectively.

For Chinese datasets, we use character-based modeling units, where WenetSpeech dataset has around 5k chars and our in-house one has around 6k chars. The English LibriSpeech dataset is trained with 1024 word-piece units, obtained with SentencePiece toolkit \cite{kudo2018sentencepiece}. Speech data in our experiments are transformed into 80-dimensional raw FBank features, CMVN \cite{viikki1998cepstral} is applied after that. In the experiments, the encoder of RNN-T is Conformer \cite{gulati2020conformer} with $1/4$ subsampling inserted before. The prediction network is 1-layer unidirectional LSTM. And the joint network is standard fully-connected layers with $\tanh(\cdot)$ activation.
The design of RNN-T components follows \cite{gulati2020conformer}, but some of the hyperparameters (e.g., hidden dimensions and number of layers of encoders) may differ in experiments. The RNN-Ts are trained with Adam optimizer and the Transformer learning rate scheduler \cite{vaswani2017attention}. At the convergence, we take best 10 of the checkpoint models and do model averaging to get the one for further evaluation.

For leveraging the performance and decoding speed, we run the decoding in monotonic topology, basically following \cite{tripathi2019monotonic}, with beam size limited to 128. The LM integration methods are evaluated in the manner of rescoring.
Hyperparameters of all LM integration methods are tuned as described in Sec.~\ref{sec:impl of lm interpolation}. Following \cite{mcdermott2019density}, LMs serving as the ILM in the DR method are 6-layer LSTM with 512-dimensional hidden size, trained on the transcripts. We train N-gram models with modified Kneser-Ney smoothing using the toolkit KenLM \cite{heafield2011kenlm}.

\vspace{-2mm}
\subsection{In-domain evaluation}
\vspace{-2mm}
\label{sec:in-domain}
The hyperparameters of LM integration are tuned on: \textit{dev} for WenetSpeech; \textit{dev-clean+dev-other} for LibriSpeech. We set the initial range [0, 1] and minimum interval size 0.1 for parameter tuning. Though the scales of ILMs $\lambda_0$ are suggested to take negative in \cite{mcdermott2019density,meng2021internal}, we do not add such restriction. With the range extending mechanism descried in Sec.~\ref{sec:impl of lm interpolation}, $\lambda_0$ is still possible to be negative.

The RNN-T model in WenetSpeech experiment is of around 90M parameters, while the one in LibriSpeech is Conformer-L with 120M parameters, following \cite{gulati2020conformer}.
The ELM used in LibriSpeech experiment is a Transformer LM with 87M parameters, trained on the 800M-word LibriSpeech corpus. No extra data is used, so our LibriSpeech results in Table.~\ref{tab:in-domain} are comparable to those in literature \cite{gulati2020conformer,zhou2021language}. WenetSpeech experiments take the 200M-char WenetSpeech corpus to train the ELM. The ELMs are shared for all methods on each dataset.
\begin{table}[t]
  \caption{Performance of LM integration methods evaluated on cross-domain scenarios.}
  \vspace{-2mm}
  \label{tab:cross-domain}
  \centering
  \setlength{\tabcolsep}{0.8mm}{
    \begin{tabular}{|c|c|c|c||c|c|c||c|}
      \hline
      \multirow{2}{*}{Method} & \multirow{2}{*}{$\lambda_0$} & \multirow{2}{*}{$\lambda_1$} & \multirow{2}{*}{$\beta$} & \multicolumn{4}{c|}{LibriSpeech $\rightarrow$ Tedlium-2}                                                  \\
      \cline{5-8}
                              &                              &                              &                          & dev                                                      & test          & avg.          & Rel \%         \\
      \hline
      No LM                   & -                            & -                            & -                        & 11.67                                                    & 11.41         & 11.51         & -              \\
      \hline
      SF                      & -                            & 0.625                        & 1.5                      & 10.26                                                    & 10.05         & 10.13         & 12.0           \\
      \hline
      DR                      & -0.125                       & 0.625                        & 1.5                      & 10.21                                                    & 9.85          & \textbf{9.99} & \textbf{13.2 } \\
      \hline
      ILME                    & -0.125                       & 0.5                          & 1.0                      & 10.23                                                    & 9.87          & 10.01         & 13.0           \\
      \hline
      LODR                    & -0.125                       & 0.625                        & 1.5                      & 10.25                                                    & 9.97          & 10.08         & 12.4           \\
      \hline
      \hline

      \multirow{2}{*}{Method} & \multirow{2}{*}{$\lambda_0$} & \multirow{2}{*}{$\lambda_1$} & \multirow{2}{*}{$\beta$} & \multicolumn{4}{c|}{WenetSpeech $\rightarrow$ AISHELL-1}                                                  \\
      \cline{5-8}
                              &                              &                              &                          & dev                                                      & test          & avg.          & Rel \%         \\
      \hline
      No LM                   & -                            & -                            & -                        & 6.32                                                     & 7.22          & 6.63          & -              \\
      \hline
      SF                      & -                            & 0.5                          & 1.375                    & 5.11                                                     & 5.56          & 5.26          & 20.7           \\
      \hline
      DR                      & -0.125                       & 0.5                          & 1.375                    & 5.10                                                     & 5.65          & 5.28          & 20.4           \\
      \hline
      ILME                    & -0.125                       & 0.5                          & 1.125                    & 4.99                                                     & 5.55          & 5.18          & 21.9           \\
      \hline
      LODR                    & -0.375                       & 0.625                        & 0.375                    & 4.76                                                     & 5.33          & \textbf{4.95} & \textbf{25.3}  \\
      \hline
      \hline

      \multirow{2}{*}{Method} & \multirow{2}{*}{$\lambda_0$} & \multirow{2}{*}{$\lambda_1$} & \multirow{2}{*}{$\beta$} & \multicolumn{4}{c|}{Tasi $\rightarrow$ TV-news}                                                           \\
      \cline{5-8}
                              &                              &                              &                          & \multicolumn{3}{c||}{tv-news test}                       & Rel \%                                         \\
      \hline
      No LM                   & -                            & -                            & -                        & \multicolumn{3}{c||}{13.41}                              & -                                              \\
      \hline
      SF                      & -                            & 0.25                         & 2.125                    & \multicolumn{3}{c||}{11.67}                              & 13.0                                           \\
      \hline
      DR                      & -0.125                       & 0.25                         & 1.625                    & \multicolumn{3}{c||}{11.61}                              & 13.4                                           \\
      \hline
      ILME                    & -0.125                       & 0.25                         & 1.375                    & \multicolumn{3}{c||}{11.46}                              & 14.5                                           \\
      \hline
      LODR                    & -0.125                       & 0.25                         & 1.25                     & \multicolumn{3}{c||}{\textbf{11.44}}                     & \textbf{14.7}                                  \\
      \hline
    \end{tabular}}

  \vspace{-5.5mm}
\end{table}

As Table.~\ref{tab:in-domain} shows, all methods bring significant improvement over the baseline (standalone RNN-T without LM). The three methods with ILM estimated (DR, ILME and LODR) are of close performance, and all outperform the SF method.

Note that all $\lambda_0$ in the Table.~\ref{tab:in-domain} are 0 or -0.125. Considering that the minimum interval size in hyperparameter searching is 0.1, this could reveal that for in-domain evaluation, the subtracted part might be unimportant. This somehow shows the SF method is sufficiently good for in-domain test.

Among the methods estimating ILMs, in LibriSpeech experiment, DR, ILME and LODR perform considerably close on average; in WenetSpeech experiment, our LODR method gains 1.2\% relative CER reduction over DR and slightly outperforms ILME.

It is also shown in the Table.~\ref{tab:in-domain} that the estimated ILM of DR simulate the transcript best (with the smallest perplexities on transcripts, in both datasets), but the evaluation performance does not consistently surpass the ILME and LODR, which further verifies our analysis in Sec.~\ref{sec:design of low-order ilm}.

\vspace{-1mm}
\subsection{Cross-domain evaluation}
\label{sec:cross-domain}
In cross-domain evaluation, we follow most of the settings in in-domain test, except that the evaluation is on a new domain and the ELM is trained on the target domain corpus.

Table.~\ref{tab:cross-domain} shows the performance of the integration methods in domain adaptation. When adapting the RNN-T trained on LibriSpeech to Tedlium-2, DR gets the best WER on both dev and test sets. DR, ILME and LODR obtain 1.4\%, 1.2\% and 0.5\% relative WER reduction on the Tedlium-2 \textit{test} compared to SF.
This relative reduction is much smaller than the literature results \cite{zhou2021language}, which reports the DR gains 8.5\% (16.4 $\rightarrow$ 15.0) relative WER reduction and ILME gains 12.2\% (16.4 $\rightarrow$ 14.4) compared to SF. We argue that, this is probably due to our significantly lower baseline (No LM: 11.41 vs. 20.3 \cite{zhou2021language}).

In the adaptation from WenetSpeech to AISHELL-1, our proposed LODR method obtains 6.2\% and 4.4\% relative CER reduction over DR and ILME methods, respectively. Overfitting of weights tuning is observed in the adaptation from WenetSpeech to AISHELL-1: compared to SF, the DR method obtains lower CER on dev set, but performs worse on test set with 1.6\% relative CER increase; ILME gains 2.2\% relative CER reduction on dev set, while only 0.2\% on test. In contrast, our LODR method obtains 4.6\% and 4.1\% relative CER reduction on the two sets over SF, showing that LODR is a very promising method for cross-domain LM adaptation.
When adapted to TV-news from our in-house Tasi dataset, LODR also outperforms all SF, DR and ILME methods.



\vspace{-1.5mm}
\subsection{Discussion}
\vspace{-1mm}

\label{sec:discussion of LODR}
As we explained in Sec.~\ref{sec:design of low-order ilm}, LODR is driven by estimating the ILM for RNN-T with the cognition that the ILM is indeed a low-order one.
\begin{table}[t]
  \caption{Comparisons of the low-order LM trained on transcripts and external corpus.}
  \label{tab:wlm source differ}
  \vspace{-1.5mm}
  \centering
  \setlength{\tabcolsep}{2mm}{
    \begin{tabular}{|c|c|c|c||c|c|}
      \hline
      Data            & $\lambda_0$ & $\lambda_1$ & $\beta$ & dev   & test \\
      \hline
      \multicolumn{6}{|c|}{LibriSpeech $\rightarrow$ Tedlium-2}            \\
      \hline
      transcripts     & -0.125      & 0.625       & 1.5     & 10.25 & 9.97 \\
      \hline
      external corpus & -0.125      & 0.625       & 1.5     & 10.22 & 9.95 \\
      \hline
      \multicolumn{6}{|c|}{WenetSpeech $\rightarrow$ AISHELL-1}            \\
      \hline
      transcripts     & -0.375      & 0.625       & 0.375   & 4.76  & 5.33 \\
      \hline
      external corpus & -0.375      & 0.625       & 0.375   & 4.75  & 5.33 \\
      \hline
    \end{tabular}}
  \vspace{-6mm}
\end{table}

To further investigate whether LODR really benefit from the transcript information, we do an ablation study that train the low-order LM from an external corpus. Here the external corpus is randomly taken from the ELMs training corpus described in Sec.~\ref{sec:in-domain}, keeping the size the same as the transcripts. It is interesting that the low-order LMs trained on transcript and external corpus have consistent performance in cross-domain adaptation on both Chinese and English tasks.
An intuitive interpretation is that, under the low-order bi-gram modeling and with most of the bi-grams pruned, the LM only learns very basic statistics about the language itself, which are, to some extent, consistent across corpus.

Reviewing the results in Table.~\ref{tab:in-domain} and Table.~\ref{tab:cross-domain}, it seems LODR performs consistently better in Chinese tasks, but not as in English. This is possibly due to the smaller number of modeling units in English tasks, which may require more than 20k bi-grams in the pruned low-order LM. Note that the number ``20k'' is an empirically set value, may not be optimal.

\vspace{-1mm}
\section{Conclusion}
In this work, we first review existing LM integration methods, including SF, DR and ILME, in the common RNN-T framework.
As recent studies suggest that RNN-T only learns some low-order LM information, we hypothesize that the ILM used in the original DR method is unduly strong and may deteriorate the performance. A low-order density ratio method (LODR) is proposed by training a low-order weak ILM for DR.
Extensive empirical experiments are conducted on both in-domain and cross-domain scenarios on English LibriSpeech \& Tedlium-2 and Chinese WenetSpeech \& AISHELL-1 datasets. 
It is shown in this study that our proposed LODR consistently outperforms the SF, and performs better than the original DR in most tests with less extra parameters introduced. As compared to ILME, our LODR method has close performance and avoids feeding the labels to the text encoder twice.
This verifies our hypothesis and deepens our understanding of the ILM in RNN-T, and would be helpful for future work to further advance LM integration for RNN-T.


\clearpage
\bibliographystyle{IEEEtran}
\bibliography{mybib}

\end{document}